\documentclass{jfm}
\usepackage{graphicx,amsmath,amssymb}
\usepackage{natbib}
\usepackage{color}

\ifCUPmtlplainloaded \else
  \checkfont{eurm10}
  \iffontfound
    \IfFileExists{upmath.sty}
      {\typeout{^^JFound AMS Euler Roman fonts on the system,
                   using the 'upmath' package.^^J}%
       \usepackage{upmath}}
      {\typeout{^^JFound AMS Euler Roman fonts on the system, but you
                   dont seem to have the}%
       \typeout{'upmath' package installed. JFM.cls can take advantage
                 of these fonts,^^Jif you use 'upmath' package.^^J}%
      }
  \else
  \fi
\fi

\ifCUPmtlplainloaded \else
  \checkfont{msam10}
  \iffontfound
    \IfFileExists{amssymb.sty}
      {\typeout{^^JFound AMS Symbol fonts on the system, using the
                'amssymb' package.^^J}%
       \usepackage{amssymb}%
         
         \let\geq=\geqslant
      }{}
  \fi
\fi

\ifCUPmtlplainloaded \else
  \IfFileExists{amsbsy.sty}
    {\typeout{^^JFound the 'amsbsy' package on the system, using it.^^J}%
     \usepackage{amsbsy}}
    {}
\fi

\newsavebox{\astrutbox}
\sbox{\astrutbox}{\rule[-5pt]{0pt}{20pt}}

\newcommand{\red}[1]{\textcolor{black}{#1}}
\newcommand{\blue}[1]{\textcolor{black}{#1}}
\newcommand{\green}[1]{\textcolor{black}{#1}}

\newcommand{\textin}[1]{\mbox{\scriptsize{#1}}}

\title[]{Global stability analysis of axisymmetric liquid-liquid flow focusing}
\author[M. G. Cabezas et al.]{M. G. Cabezas$^{1}$, N. Rebollo-Mu\~noz$^{1}$, M. Rubio$^{1}$, M. A. Herrada$^{2}$,\\ J. M. Montanero$^{1}$}

\affiliation{$^1$Departamento de Ingenier\'{\i}a Mec\'anica, Energ\'etica y de los Materiales and\\
Instituto de Computaci\'on Cient\'{\i}fica Avanzada (ICCAEx),\\
Universidad de Extremadura, E-06071 Badajoz, Spain\\[\affilskip]
$^2$Departamento de Ingenier\'{\i}a Aeroespacial y Mec\'anica de Fluidos,\\
Universidad de Sevilla, E-41092 Sevilla, Spain}

\begin{document}

\maketitle

\begin{abstract}
We analyze both numerically and experimentally the stability of the steady jetting tip streaming produced by focusing a liquid stream with another liquid current when they coflow through the orifice of an axisymmetric nozzle. We calculate the global eigenmodes characterizing the response of this configuration to small-amplitude perturbations. In this way, the critical conditions leading to the instability of the steady jetting tip streaming are determined. The \blue{unstable perturbations} are classified according to their oscillatory character and to the region where they are originated (convective and absolute instability). We derive and explain in terms of the velocity field a simple scaling law to predict the diameter of the emitted jet. The numerical stability limits are compared with experimental results finding reasonable agreement. \red{The experiments confirm the existence of the two instability mechanisms predicted by the global stability analysis.}
\end{abstract}

\section{Introduction}
\label{s1}

The axisymmetric liquid-liquid flow focusing configuration has been frequently studied because of its very diverse applications \citep{MG20}. For example, a flow cytometer can be built  by focusing a liquid stream across coaxial converging nozzles \citep{LHKHHL01}. In the pioneering work of \citet{ULLKSW05} and subsequent experimental studies \citep{TGWW05,SSHW10,NVM17}, axisymmetric liquid-liquid flow focusing was proposed for encapsulation and release of different actives. Monodisperse cell-encapsulating microgel beads \citep{TMT10} and biodegradable Poly(Lactic Acid) particles \citep{VHDSW12} have been fabricated using flow focusing. Multiple-emulsions are ideal microreactors or fine templates for synthesizing advanced particles \citep{WWH11}. These emulsions have been produced with a high degree of monodispersity with the double flow focusing configuration \citep{CLSSW11,CWZ15}. The use of liquid-liquid flow focusing to produce emulsions has been reviewed by \citet{GDM11}. \citet{WYLXZSX17} have proposed multiplex coaxial flow focusing for single-step fabrication of multicompartment Janus microcapsules. PDMS microcapsules with tunable elastic properties  \citep{NAMMCD17} and multi-compartment polymeric microcapsules \citep{ZWHXZYDSX18} can be formed with axisymmetric liquid-liquid flow focusing too. Using a similar configuration, the so-called ``impinging flow-focusing", \citet{WLDCH17} produced monodispersed emulsions at large frequencies with very small diameters. 

The axisymmetric liquid-liquid flow focusing configuration can operate in the so-called tip streaming mode \citep{MG20}. In tip streaming, a droplet or meniscus attached to a feeding capillary ejects tiny droplets that are much smaller than any characteristic length of the fluid configuration. This effect is produced by the collaboration between normal and shear stresses acting on the interface. The normal stress stretches the apex of the droplet/meniscus against the action of the surface tension force. In this way, the apex adopts the shape of a minuscule fluid nozzle which constitutes the gate for the liquid ejection. This ejection is essentially driven by the shear stress at the interface, which powers a thin fluid layer on the internal side of the interface. If the kinetic energy of this layer is sufficiently large to overcome the resistance offered by both the surface tension and viscosity, the fluid ejection takes place. This ejection undergoes an initial transient phase that arises right after the appearance of the driving stresses. In this transient phase, a fluid thread, long compared with its diameter, is formed. The thread formation is followed by the quasi-steady ejection of a liquid microjet (steady jetting tip streaming, SJTS), which eventually breaks up into droplets due to the capillary instability. The SJTS mode keeps running if the expelled volume is properly replaced by injecting liquid into the tapering droplet/meniscus. The diameter of the emitted jet can be reduced by decreasing the injected flow rate. If the rest of the parameter conditions are fixed, there is a minimum value of the flow rate below which SJTS becomes unstable. 

\green{Liquid-liquid SJTS can also be produced in configurations similar to flow focusing, such as the coflowing configuration \citep{SB06,MCG09,RSG13}, where the effect produced by the discharge orifice is not present, or the selective withdrawal \citep{CLHMN01} and confined selective withdrawal \citep{HCGLG19} configurations, where the focusing phenomenon is caused by a cylindrical capillary located in front of a liquid film and meniscus, respectively.} 

Liquid-liquid flow-focusing devices produce a strong focusing effect in front of the discharge orifice owing to the collaboration between the drop of hydrostatic pressure and outer viscous stresses in that region. This collaboration favors the transition to SJTS. In fact, this regime can be obtained in flow focusing for outer stream velocities much smaller than those necessary in an equivalent coflowing configuration \citep{LWZC17,WLZC17}. In the flow focusing region, the drop of hydrostatic pressure and the viscous stress scale as $D^{-4}$ and $D^{-2}$, respectively. For this reason, the orifice diameter $D$ considerably affects the jet formation. The focusing effect can be enhanced by appropriately modifying the focusing geometry. The so-called ``opposed flow focusing"\ configuration \citep{DMFESR18} seems to lead to a second-order transition that enables a sharp reduction of the jet radius down to vanishing scales.

Axisymmetric liquid-liquid flow focusing applied to complex fluids has also received attention. For instance, numerical simulations show that viscoelasticity delays the transition from dripping to SJTS \citep{NIM16}. When soluble surfactants are present, tip streaming essentially occurs when the mass of surfactant adsorbed to the interface is that needed to maintain the interfacial conical shape \citep{MWA12}. The outer velocity field in axisymmetric surfactant-mediated flow focusing has been described by combining an imposed uniaxial extension flow at infinity with two transverse, coaxial, annular baffles placed symmetrically to either side of the drop \citep{WBSW18}. SJTS results from the interfacial reduction caused by the soluble surfactant and the focusing effect produced by the baffles.

The stability analysis of the steady tip streaming produced by axisymmetric liquid-liquid flow focusing has received very little attention. In their pioneering work, \citet{GR06} studied the jetting-to-dripping transition when the liquid stream is focused by another liquid current across a circular orifice. They compared the critical inner-to-outer flow rate ratios in the experiments with those leading to the convective-to-absolute instability transition in the jet. However, this comparison is pertinent only if the instability is originated in that part of the fluid domain. When the source of instability is localized in the tapering meniscus, the jet local stability analysis cannot predict the critical parameter conditions. In fact, the experimental stability limit exhibited an ``elbow"\ in the plane defined by the jet's Reynolds and Weber numbers which cannot be described from the convective-to-absolute instability transition in the emitted jet \citep{MG08a}. \citet{MDS18} have recently studied both experimentally and theoretically the stability of the axisymmetric liquid-liquid flow focusing configuration. They have successfully explained their experimental observations and direct numerical simulations by distinguishing the instability originated in the emitted jet from that localized in the tapering meniscus.

The minimum inner flow rate to get SJTS inside a converging-diverging nozzle was reduced up to two orders of magnitude by injecting the focused liquid through a hypodermic needle \citep{ARMGV13}. The essential idea was to replace the tapering meniscus hanging on the feeding capillary with a film sliding over the hypodermic needle tip. The importance of the stability of the complex flow pattern arising in the tapering meniscus of flow focusing has been pointed out for the gaseous configuration too \citep{VMHG10,MRHG11}.

The calculation of the linear global modes \citep{T11} is an adequate tool to predict the instability of SJTS. The idea is to assume that a long jet tapers from the liquid meniscus. Then, we interrogate this basic flow about its response to small-amplitude perturbations \citep{SB05a,TLS12,GSC14}. SJTS becomes unstable if the largest growth rate of the eigenfrequency spectrum is positive. In this case, the system evolution is asymptotically (i.e., for sufficiently large times) dominated by the corresponding mode whose growth leads to either the interruption of the ejection or to self-sustained oscillations. The complexity of the global stability analysis probably explains why it has not been applied to many SJTS configurations \citep{MG20}, including axisymmetric liquid-liquid flow focusing.

The global stability analysis of SJTS entails an additional difficulty not present in confined capillary systems. With very few exceptions \citep{HAFSW09, CHM19b}, an infinite capillary jet is intrinsically unstable; i.e., perturbations with sufficiently large wavelengths grow over time due to the action of the surface tension independently of the parameter conditions. This implies that SJTS becomes unstable if the fluid domain considered in the analysis is sufficiently extended in the downstream direction. Therefore, the stability or instability of the SJTS realization cannot be regarded as an intrinsic property of this flow in a strict sense, but depends on the cutoff imposed downstream in the numerical analysis. Therefore, there is inevitably a certain degree of arbitrariness in the determination of SJTS stability. \blue{This problem is typically addressed by imposing two conditions to the downstream cutoff: (i) the numerical domain must be sufficiently large to contain a very long jet compared to its diameter, and (ii) the critical value of the control parameter must not significantly change when the cutoff is considerably varied. This last condition implies that the jet length (the so-called intact region length) must experience a sharp reduction from values much larger than the selected cut-off length to values much smaller than that parameter in the vicinity of the jetting-to-dripping transition. In this way, the threshold for the jetting-to-dripping transition hardly depends on the cutoff length and the theoretical prediction can be compared with experiments.} 

The above-described difficulty to conduct the global stability analysis of SJTS has its counterpart in the experiments too. There is also an important degree of arbitrariness in identifying an experimental realization either as jetting or as dripping. The existence of the Rayleigh capillary breakup \citep{R79a} can be regarded as a criterion to ensure that a given experimental realization corresponds to jetting. However, this breakup mechanism can be combined with, and progressively replaced by, the end-pinching mechanism \citep{SBL86} as the jet shortens. On many occasions, the breakup procedure cannot be easily distinguished from dripping. 

\blue{It is frequently believed that (asymptotic) global stability is a sufficient condition for the base flow to be linearly stable. However, the short-term dynamics of the system can be the result of a ``constructive interference" of stable global modes, which can lead to a bifurcation before those modes are damped out \citep{LCC02,S07}. In other words, the superposition of decaying small-amplitude perturbations can destabilize the flow before those perturbations disappear, which prevents the system from reaching SJTS.} \citet{CHGM17} and \citet{PRHGM18} have recently obtained the stability limits of SJTS in gaseous flow focusing and the cone-jet mode of electrospray, respectively. The theoretical predictions calculated from the (asymptotic) global stability analysis agreed remarkably well with the experimental data for all the cases of electrospray analyzed \citep{PRHGM18}. However, SJTS became unstable for flow rates larger than those predicted by the stability analysis in gaseous flow focusing for small applied pressure drops \citep{CHGM17}. This instability is caused by the short-term superposition of the eigenmodes \citep{LCC02,S07}. A natural question is which of these two possible scenarios the axisymmetric liquid-liquid flow focusing configuration corresponds to.

In this paper, we will examine the stability of the SJTS produced by axisymmetric liquid-liquid flow focusing both numerically and experimentally. We will calculate both the base flow and its linear eigenmodes to determine the critical parameter conditions at which SJTS becomes unstable. Both oscillatory and non-oscillatory \blue{unstable perturbations} will be identified. We will show that these perturbations can be originated in the tapering meniscus (absolute instability) or beyond the discharge orifice (convective instability) depending on the values of the viscosity and flow rate ratios. The jet diameter will be obtained at the minimum inner flow rate stability limit. We will derive a simple scaling law for that quantity. The comparison with experiments will show good agreement over the whole range of flow rate ratios analyzed. \red{The existence of the two instability mechanisms predicted by the global stability analysis will be confirmed by the experiments.}

\section{Axisymmetric liquid-liquid flow focusing}
\label{s2}

In the axisymmetric liquid-liquid flow focusing configuration considered in this paper, a liquid of density $\rho_i$ and viscosity $\mu_i$ is injected through a cylindrical feeding capillary of radius $R_c$ at a constant flow rate $Q_i$. The feeding capillary is located inside a converging-diverging nozzle whose inner shape is defined by the function $\hat{S}(\hat{z})$, which measures the distance of the inner contour to the symmetry axis $\hat{z}$. The most important parameter of this contour is the neck diameter $\hat{D}$. The distance between the capillary end and the nozzle neck is $\hat{H}$. A liquid stream of density $\rho_o$ and viscosity $\mu_o$ flows through the nozzle at a constant flow rate $Q_o$. The interfacial tension of the interface between the two immiscible liquids is $\sigma$. In SJTS, an axisymmetric meniscus attached to the edge of the capillary end emits a microjet from its tip. 

The emitted jet crosses the nozzle orifice coflowing with the outer liquid stream. To produce the flow focusing effect, the focusing (outer) stream speed in the nozzle orifice must considerably exceed that of the focused (inner) current, which makes the jet accelerate in that region. Both the focused and focusing streams discharge into the liquid bath. The jet acceleration continues downstream as long as the speed of the outer stream exceeds that of the jet. The viscous drag force exerted by the outer bath on the focusing current makes the latter slow down. Viscous radial diffusion of momentum causes the jet deceleration far away from the discharge orifice, which makes the jet radius increase. We define the jet radius $R_j$ as the minimum radius of the focused current within the fluid domain analyzed. 

\green{The dimensionless governing parameters are the density and viscosity ratios, $\rho\equiv\rho_o/\rho_i$ and $\mu=\mu_o/\mu_i$, the flow rate ratio $Q\equiv Q_o/Q_i$, and the Capillary and Reynolds numbers, Ca$=\mu_i Q_i/(\pi R_c^2 \sigma)$ and Re$\equiv \rho_i Q_i/(\pi R_c\mu_i)$, defined in terms of the inner liquid properties, the interfacial tension, and the inner flow rate. As occurs in the coflowing configuration \citep{SB06}, the Reynolds number is expected to play a secondary role because of the small effect of inertia. The formulation of the problem is completed by fixing the dimensionless nozzle shape $S(z)$ ($S\equiv \hat{S}/R_c$, $z\equiv \hat{z}/R_c$) and the capillary position $H\equiv \hat{H}/R_c$.}

\section{Governing equations}
\label{s3}

In this section, all the variables are made dimensionless with the capillary radius $R_c$, the mean inlet velocity $v_c=Q_i/(\pi R_c^2)$, and the liquid viscosity $\mu_i$, which yields the characteristic time and stress $t_c=\pi R_c^3/Q_i$ and $p_c=\mu_i v_c/R_c$, respectively. The dimensionless, axisymmetric, incompressible Navier-Stokes equations for the velocity $\mathbf{v}^{(k)}(r,z;t)$ and pressure $p^{(k)}(r,z;t)$ fields are
\begin{eqnarray}
\left(ru^{(k)}\right)_r+rw^{(k)}_z &=&0, \label{basic1}\\
\rho^{\delta_{ko}} \text{Re}\, \left(u^{(k)}_t + u^{(k)} u^{(k)}_r+ w^{(k)} u^{(k)}_z\right)&=&-p^{(k)}_r+\mu^{\delta_{ko}}
\left(u^{(k)}_{rr}+(u^{(k)}/r)_r+u^{(k)}_{zz}\right), \label{basic2}\\
\rho^{\delta_{ko}} \text{Re}\, \left(w^{(k)}_t + u^{(k)} w^{(k)}_r+ w^{(k)}w^{(k)}_z\right) &=&-p^{(k)}_z+\mu^{\delta_{ko}\,} \left(w^{(k)}_{rr}+w^{(k)}_r/r+w^{(k)}_{zz}\right),\label{basic3}
\end{eqnarray}
where $t$ is the time, $r$ ($z$) is the radial (axial) coordinate, $u^{(k)}$ ($w^{(k)}$) is the radial (axial) velocity component, and $\delta_{ij}$ is the Kronecker delta. In the above equations and henceforth, the superscripts $k=i$ and $o$ refer to the inner and outer phases, respectively, while the subscripts $t$, $r$, and $z$ denote the partial derivatives with respect to the corresponding variables. The action of the gravitational field has been neglected due to the smallness of the fluid configuration.

Taking into account the kinematic compatibility and equilibrium of tangential and normal stresses at the interface $r=F(z,t)$, one gets the following equations:
\begin{equation}
F_t+F_z w^{(i)}-u^{(i)}=F_t+F_z w^{(o)}-u^{(o)}=0,\label{int1}
\end{equation}
\begin{equation}
(1-F_z^{2})(w^{(i)}_r+u^{(i)}_z)+2F_z(u^{(i)}_r-w^{(i)}_z)=\mu[(1-F_z^{2})(w^{(o)}_r+u^{(o)}_z)+2F_z(u^{(o)}_r-w^{(o)}_z)], \label{int2}
\end{equation}
\begin{eqnarray}
&&p^{(i)}+\text{Ca}^{-1}\frac{FF_{zz}-1-F_z^{2}}{F(1+F_z^{2})^{3/2}}-\frac{2[u^{(i)}_r-F_z(w^{(i)}_r+u^{(i)}_z)+F_z^{2}w^{(i)}_z]}{1+F_z^{2}}=
\nonumber\\&&
p^{(o)}-\frac{2\mu[u^{(o)}_r-F_z(w^{(o)}_r+u^{(o)}_z)+F_z^{2}w^{(o)}_z]}{1+F_z^{2}}.\label{int3}
\end{eqnarray}

\begin{figure}
\begin{center}
\resizebox{0.75\textwidth}{!}{\includegraphics{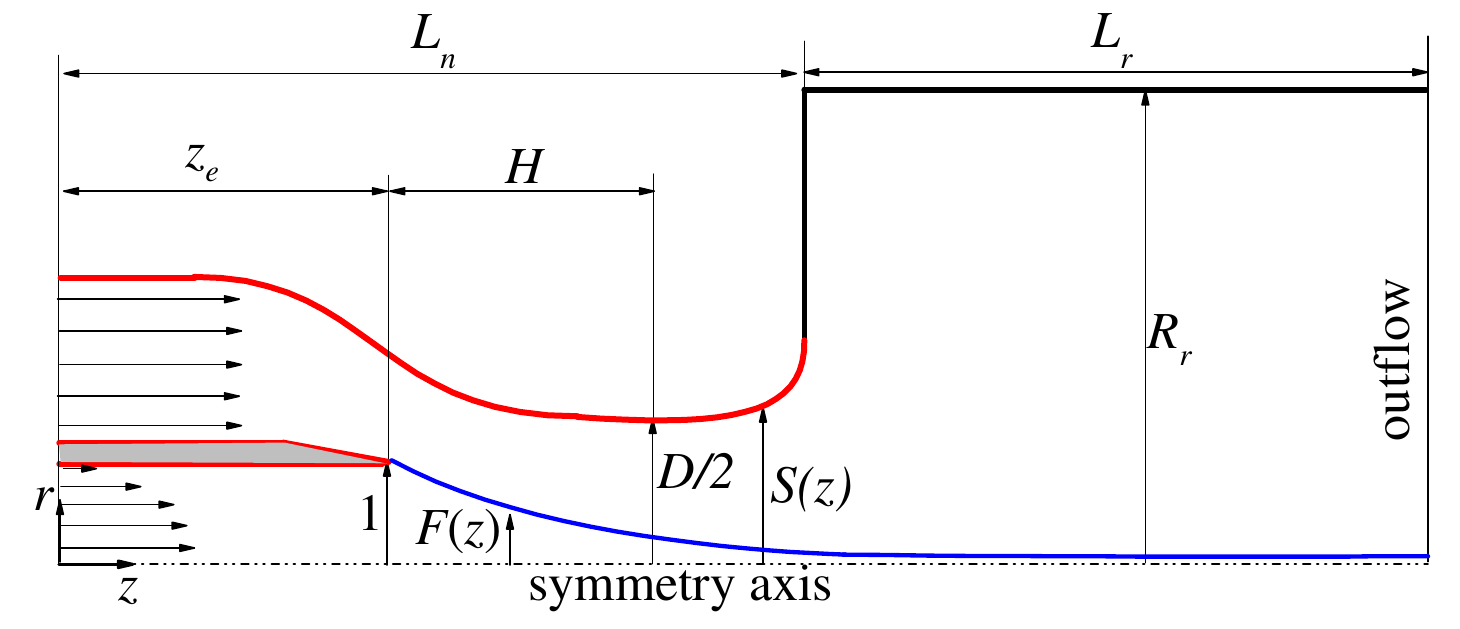}}
\end{center}
\caption{Sketch of the computational domain.}
\label{numer0}
\end{figure}

The Navier-Stokes equations are integrated in the numerical domain sketched in Fig.\ \ref{numer0}. The boundary of the domain is split into two parts. The red lines correspond to the digitized contours of the nozzle $S(z)$ and the sharpened feeding capillary used in the experiments, while the black lines have been added to close the numerical domain. The nozzle and capillary lengths in the domain are $L_n=15$ and $z_e=7.5$, respectively, the neck diameter is $D=1.98$, the distance from the needle end to nozzle neck is $H=3.8$, and the outer bath radius is $R_r=7$. The value of the cutoff length $L_r$ will be discussed below. The feeding capillary end is assumed to be infinitely thin to facilitate the numerical calculations, which may constitute a significant difference with respect to the experiments. \green{It should be noted that the simulation of blunt or flat capillary ends requires modeling the triple contact line dynamics and introducing a numerical subdomain because the mapping (see Sec.\ \ref{s4}) would be multivalued on that surface. This considerably complicates the numerical method.}

At the inlet section $z=0$, we impose a uniform axial velocity and a parabolic profile for the outer and inner fluids, respectively. The non-slip boundary condition is prescribed at the solid walls. The free surface shape is obtained as part of the solution by considering the anchorage condition $F=1$ of the triple contact line at the edge of the feeding capillary. We impose the standard regularity conditions $u^{(i)}=w^{(i)}_r =0$ at the symmetry axis, and the outflow conditions $u^{(k)}_z=w^{(k)}_z=F_z=0$ at the right-hand end of the computational domain.

The linear global modes are calculating by assuming the temporal dependence 
\begin{equation}
U(r,z;t)=U_0(r,z)+\varepsilon\delta U(r,z)\, e^{-i\omega t} \quad (\varepsilon\ll 1),
\end{equation}
where $U(r,z;t)$ represents any hydrodynamic quantity, $U_0(r,z)$ and $\delta U(r,z)$ stand for the base (steady) solution and the spatial dependence of the eigenmode, respectively, while $\omega=\omega_r+i\omega_i$ is the eigenfrequency. Both the eigenmodes $\delta U$ and the corresponding  eigenfrequencies $\omega$ are obtained as a function of the governing parameters. The dominant eigenmode is that with the largest growth rate $\omega_i$. If that growth rate is positive, the base flow is asymptotically unstable \citep{T11}. 

\section{Numerical method}
\label{s4}

\begin{figure}
\begin{center}
\resizebox{\textwidth}{!}{\includegraphics{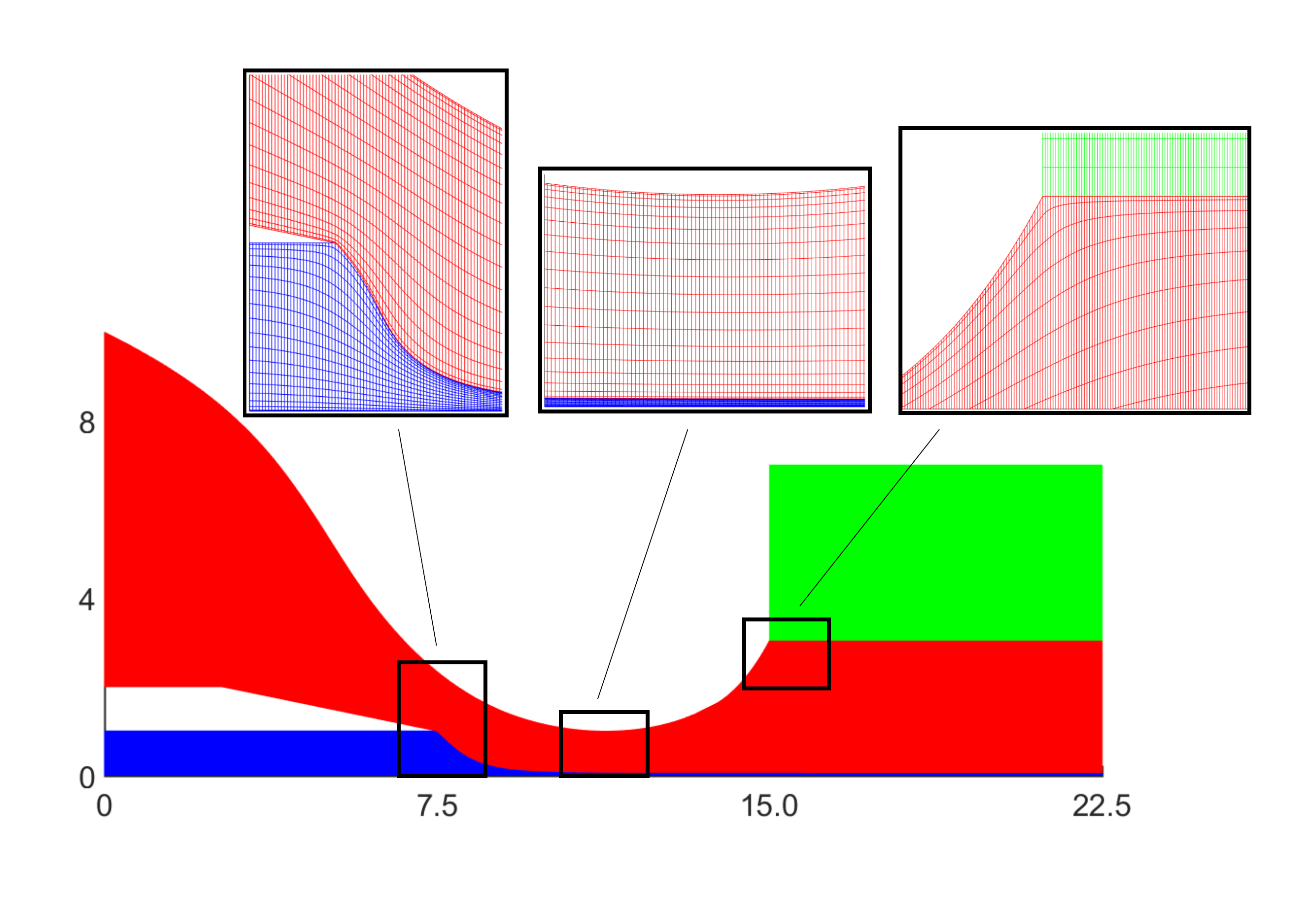}}
\end{center}
\caption{Details of the grid used in the simulations. The grid consists of three blocks corresponding to the inner liquid (blue), outer stream (red), and surrounding bath beyond the discharge orifice (green). }
\label{numer2}
\end{figure}

We used the numerical method proposed by \citet{HM16a} to solve the theoretical model described in the previous section. Here, we summarize the main characteristics of this method. The inner and outer fluid domains were mapped onto two quadrangular domains through a non-singular mapping. To smooth the effect of the corners of the feeding capillary, a quasi-elliptic transformation \citep{DT03} was applied in the blue and red regions shown in Fig.\ \ref{numer2}, while the green zone was discretized with a rigid grid. All the derivatives appearing in the governing equations were expressed in terms of $t$ and the spatial coordinates resulting from the mapping. These equations were discretized in the (mapped) radial direction with $n_{\chi}$ Chebyshev spectral collocation points in each region. We used fourth-order finite differences with $n_{\xi}$ equally spaced points to discretize the (mapped) axial direction. The results presented in this work were calculated using $n_{\chi}=19$ for each of the regions mentioned above and $n_{\xi}=1201$.

The calculation of the base flow proceeded as follows. Before running the simulation, the elements of the Jacobian ${\cal J}^{(p,q)}$ of the discretized system of equations ${\cal J}^{(p,q)}U_0^{(q)}={\cal F}^{(p)}$ for the base flow unknowns $U_0^{(q)}$ ($q=1,2,\ldots,n\times N$ stands for the values of the $n$ unknowns at the $N$ grid points) were computed via standard symbolic software at the outset. We evaluated numerically the resulting functions over the Newton-Raphson iterations, which considerably reduced the CPU time. In each of those iterations, the Jacobian ${\cal J}_0^{(p,q)}={\cal J}^{(p,q)}(U_0^{(q)})$ was evaluated for the updated value of $U_0^{(q)}$. The inverse matrix ${\cal J}_0^{-1(q,p)}$ was calculated, and the correction vector $\delta \widehat{U}_0^{(q)}=-{\cal J}_0^{-1(q,p)} {\cal F}^{(p)}$ was obtained from the functions ${\cal F}^{(p)}$ evaluated in the previous iteration. 

The above numerical procedure allows us to calculate the linear global modes too. The spatial dependence of the linear perturbation $\delta U^{(q)}$ is the solution to the generalized eigenvalue problem ${\cal J}_0^{(p,q)} \delta U^{(q)}=i\omega {\cal Q}_0^{(p,q)} \delta U^{(q)}$, where ${\cal J}_0^{(p,q)}$ is the Jacobian of the system evaluated with the base solution $U_0^{(q)}$, and ${\cal Q}_0^{(p,q)}$ accounts for the temporal dependence of the problem. This matrix was calculated with essentially the same procedure as that for ${\cal J}_0^{(p,q)}$ \citep{HM16a}.

We conducted simulations for different mesh sizes to ensure that the results did not depend on that choice. For instance, for $\mu=0.01$ and 0.1 an increase of 75\% in the number of points produced errors in the critical flow rate ratio below 0.1\%. We also verified that the results were not significantly affected by the choice of the cutoff length $L_r$. Specifically, we increased $L_r$ by 50\% and the critical flow rate ratio differed in less than 0.1\% in all the cases analyzed. All the results presented in this paper were obtained for $L_r=7.5$.

\section{Experimental method}
\label{s5}

\begin{figure}
\begin{center}
\resizebox{0.8\textwidth}{!}{\includegraphics{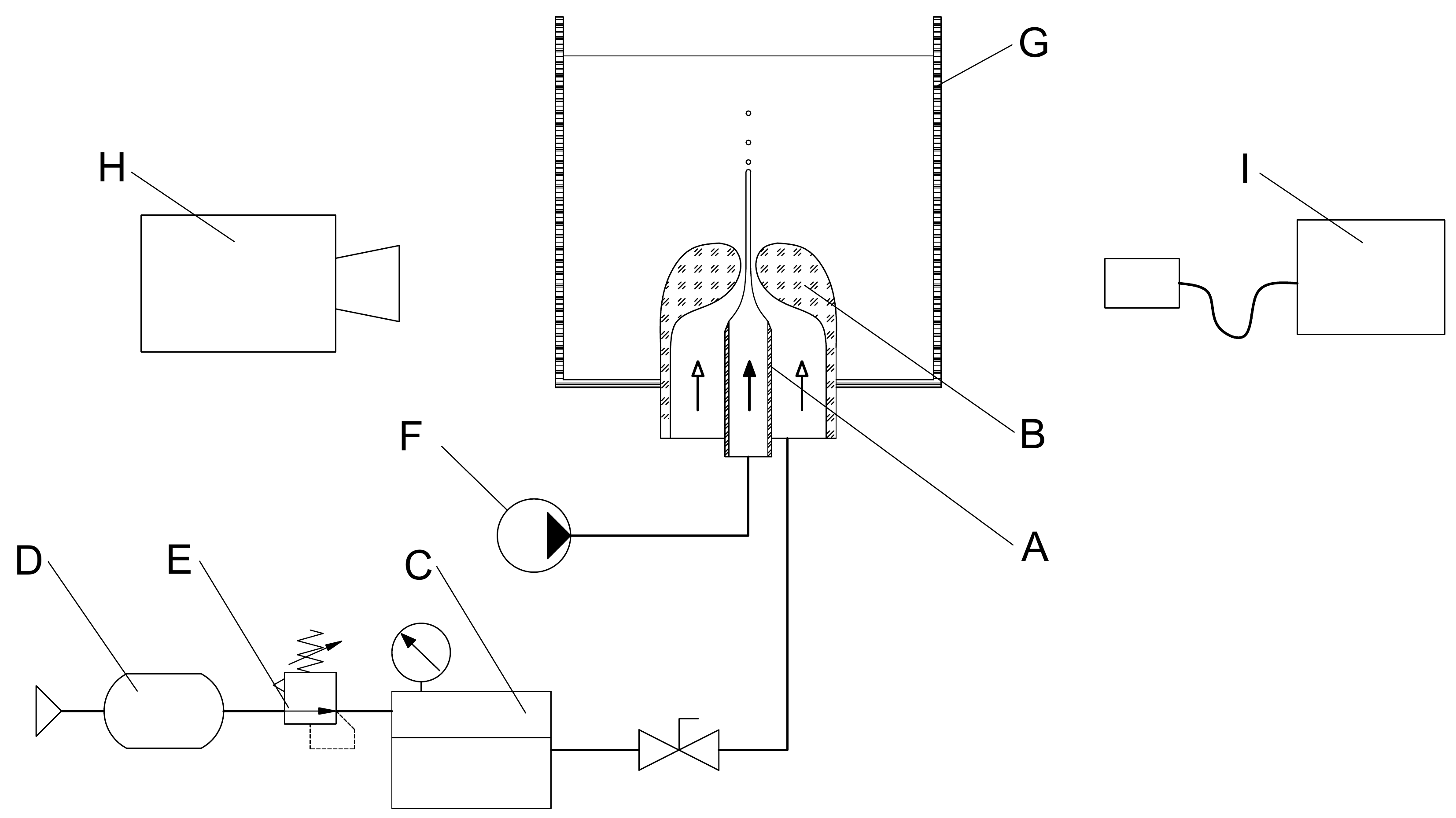}}
\end{center}
\caption{\red{Experimental setup: (A) inner feeding capillary, (B) glass nozzle, (C) pressurized reservoir, (D) compressed air tank, (E) pressure reducing valve, (F) syringe pump, (G) open cell, (H) camera, and (I) lighting system.}}
\label{Setup}
\end{figure}

\red{Figure \ref{Setup} shows a sketch of the experimental setup used in the experiments.} The flow focusing device consisted of a capillary tube (A) of inner radius $R_c=100$ $\mu$m located coaxially inside a glass converging-diverging nozzle (B). The tube end was sharpened to force the triple contact line to attach to the inner radius. The glass converging-diverging nozzles used in the experiments were fabricated with the method recently proposed by \citet{MGC19}, which produces highly axisymmetric and reproducible nozzles. The neck diameter was $\hat{D}=198$ $\mu$m. The distance between the tube end and the neck was $\hat{H}$=380 $\mu$m. The focusing liquid was injected from a reservoir (C) partially filled with the liquid and pressurized with compressed air coming from a tank (D). The pressure of the liquid free surface in the reservoir was controlled with a high-precision pressure regulating valve (E). The injection system was calibrated to determine the relationship between that pressure applied to the liquid free surface in the reservoir and the injected flow rate. The focused liquid was injected with a syringe pump (F) \red{(Legato 210, KD Scientific)}. The flow focusing device discharged both the focused and focusing streams into a transparent cell (G). We verified that the optical distortion caused by the cell was negligible.

Digital images of the fluid configuration were acquired using a \red{high-speed camera} (H) \red{(Fastcam SA5, Photron)} equipped with a set of optical lenses (a set of lenses with variable magnification from 0.75$\times$ to 5.25$\times$) with a magnification ranging from \green{??} to \green{??} $\mu$m/pixel. The camera could be displaced both horizontally and vertically using a triaxial translation stage to focus the jet. The fluid configuration was illuminated from the back with the cold white light provided by an optical fiber light source (I) \red{(KL2500 LCD, Schott)}. Images of the capillary tube were also acquired with an auxiliary CCD camera perpendicularly to the other camera to check that the capillary tube was correctly positioned inside the nozzle. All these elements were mounted on an optical table with a pneumatic antivibration isolation system to damp the vibrations coming from the building.

\begin{figure}
\begin{center}
\resizebox{0.6\textwidth}{!}{\includegraphics{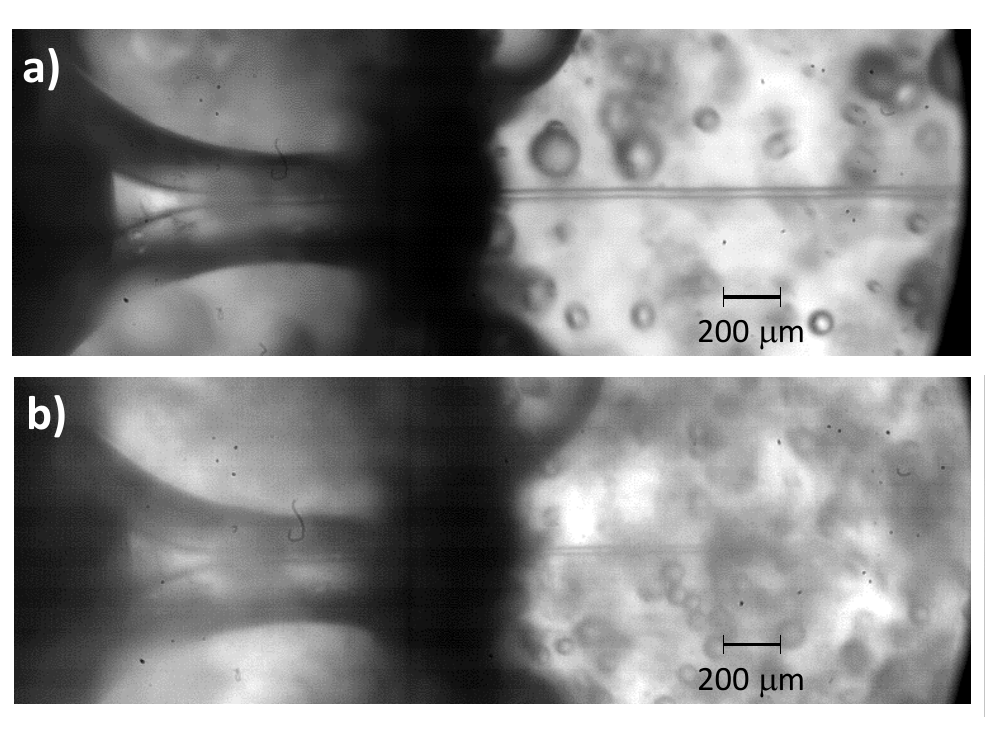}}
\end{center}
\caption{\red{Experimental images of jets of 20 cSt silicone oil produced with $Q_i$= 7 ml/h and $Q_o$= 5.5 ml/min (a), and with $Q_i$ = 2 ml/h and $Q_o$= 5.5 ml/min (b).}}
\label{F_ExpIm}
\end{figure}

In each experimental run, the two liquids were injected at flow rates sufficiently large to get SJTS. Then, the focused-liquid flow rate was progressively decreased while keeping constant that of the focusing liquid. Images of the emitted jet were acquired during this process \red{(Fig. \ref{F_ExpIm})}. \green{It should be noted that the system exhibits a hysteretic behavior, i.e. the dynamical response for a given focused-liquid flow rate depends on whether that flow rate was reached by decreasing or increasing it. However, reducing the focused-liquid flow rate is the right way to determine the stability limit if the experimental result is to be compared with the global stability analysis prediction. In the global stability analysis, we assume that a base flow has been established and wonder whether that flow withstands small-amplitude perturbations. This corresponds to what we do in our experiments by reducing the focused-liquid flow rate.}

We conducted experiments with 20 cSt and 100 cSt silicone oil \red{(Xiameter PMX-200, Dow Corning)} focused with distilled water. The physical properties of the working liquids are shown in Table \ref{tab1}. The density and viscosity values were taken from the manufacturer, while the interfacial tension between water and each silicone oil was measured with the TIFA (Theoretical Image Fitting Analysis) method \citep{CBMN04}.

\begin{table}
\begin{center}
\begin{tabular}{lcccc}
\hline Liquid &  $\rho$ (kg$\cdot$m$^{-3}$) &  $\sigma$ (mN$\cdot$m$^{-1}$) &  $\mu$ (mPa$\cdot$s)\\
\hline
20 cSt silicone oil   & 949 & 35 & 19 \\
100 cSt silicone oil  & 957 & 35 & 96\\
Water & 997 & - & 1\\
\hline
\end{tabular}
\end{center}
\caption{Properties of the working liquids at 20 $^\circ$C. The interfacial tension $\sigma$ refers to the interface between water and the corresponding oil.}
\label{tab1}
\end{table}

\section{Results}
\label{s6}

\begin{figure}
\begin{center}
\resizebox{0.6\textwidth}{!}{\includegraphics{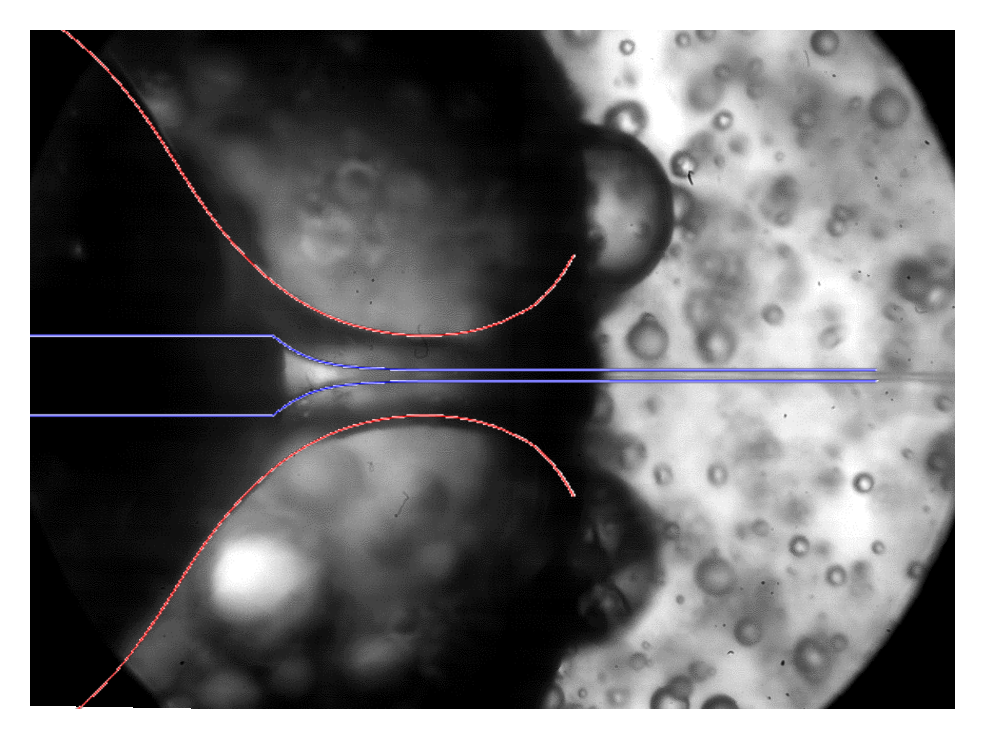}}
\end{center}
\caption{\green{Comparison between the experimental and numerical (blue line) interface contours. The inner fluid is 20 cSt silicone oil, and the flow rates are $Q_i$ = 7 ml/h, $Q_o$= 5.5 ml/min.}}
\label{valido}
\end{figure}

\green{Figure \ref{valido} compares the experimental interface contour and that calculated for the base flow in a SJTS realization. As can be observed, there is a remarkable agreement between the experimental and numerical contours. This agreement should be regarded as a prerequisite for an accurate global stability analysis because the dominant eigenmode at the stability limit may be sensitive to small errors of the base flow.} 

\begin{figure}
\begin{center}
\resizebox{0.4\textwidth}{!}{\includegraphics{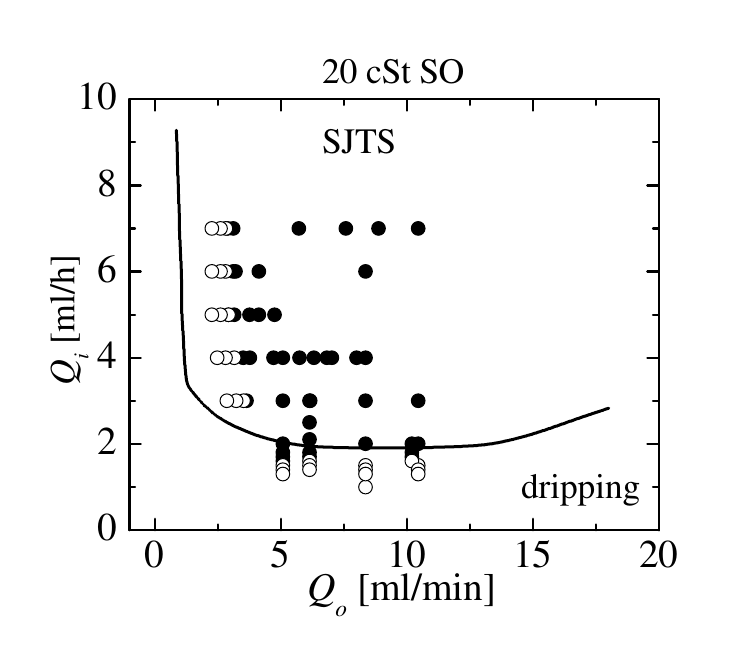}}\resizebox{0.41\textwidth}{!}{\includegraphics{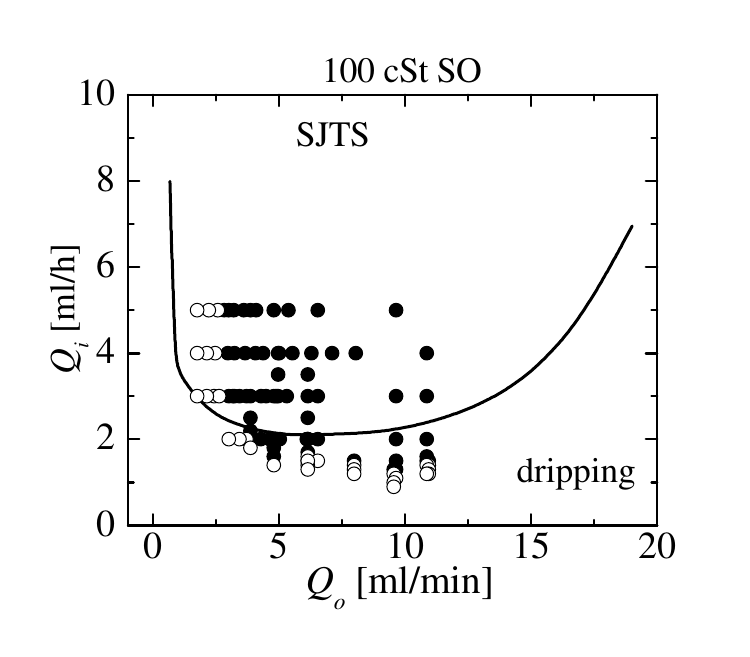}}
\end{center}
\caption{Experimental SJTS (solid symbols) and dripping (open symbols) realizations of 20 cSt and 100 cSt silicone oils focused with water. The solid line is the stability limit obtained from the global stability analysis. In the left-hand graph, the inset shows an image of a SJTS realization.}
\label{er}
\end{figure}

\begin{figure}
\begin{center}
\resizebox{0.4\textwidth}{!}{\includegraphics{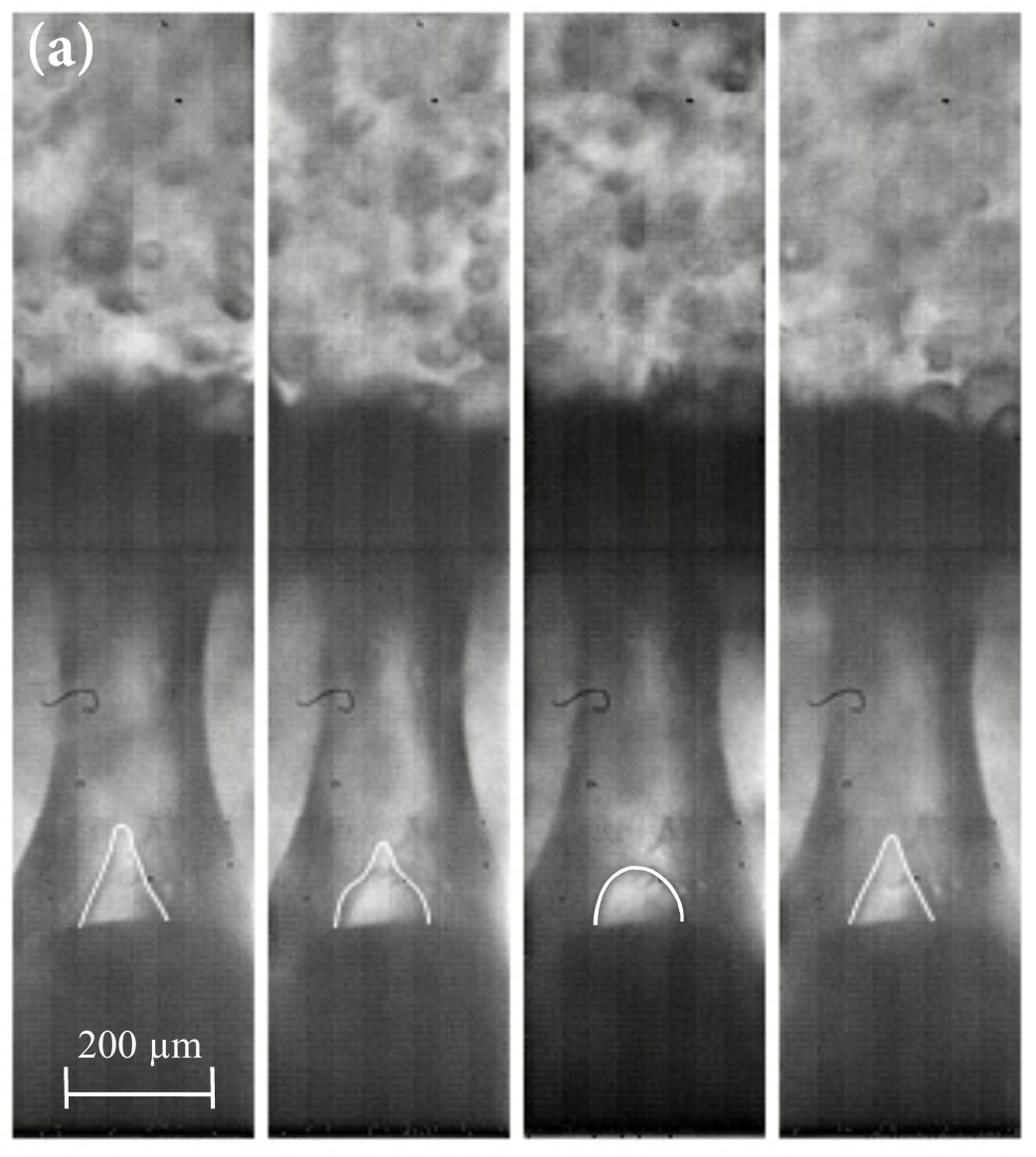}}
\resizebox{0.4\textwidth}{!}{\includegraphics{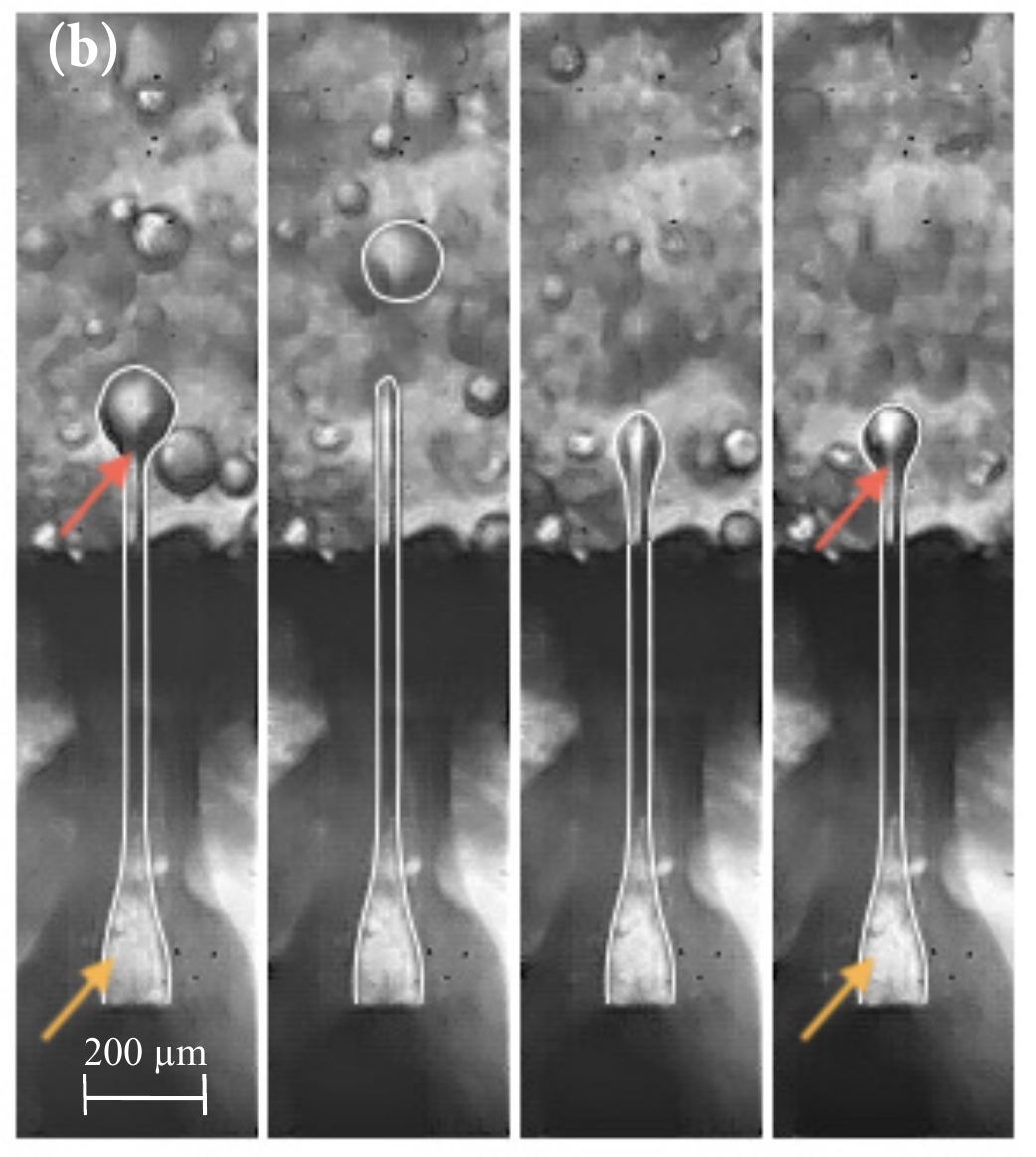}}
\end{center}
\caption{\red{Sequence of experimental images showing the meniscus instability (a) and the jet breakup (b). The white lines have been added to highlight the interface contours. In the right-hand image, the orange and red arrows indicate the quasi-steady meniscus and the droplet formed right-after the discharge orifice, respectively. The inner fluid is 20 cSt silicone oil. The flow rates are $Q_i$= 0.5 ml/h and $Q_o$= 5.5 ml/min (a), and $Q_i$= 5 ml/h and $Q_o$= 3.1 ml/min (b).}}
\label{F_inestMeniscus}
\end{figure}

Figure \ref{er} shows experimental SJTS (solid symbols) and dripping (open symbols) realizations of 20 cSt and 100 cSt silicone oils focused with water. It must be noted that in this work we regard as SJTS only the jetting realizations producing long jets as compared with the discharge orifice diameter, as also done in the global stability analysis. The solid line is the stability limit obtained from that analysis. As will be explained below, the global stability analysis predicts the existence of two instability mechanisms: the destabilization of the tapering meniscus (absolute instability) and the loss of stability of the emitted jet beyond the discharge orifice (convective instability). The horizontal and vertical branches of the stability limit curve in Fig.\ \ref{er} correspond to the former and latter mechanisms, respectively. The experiments show the validity of the above prediction. In fact, we verified that the dripping observed below the horizontal branch is caused by the intermittent ejection of liquid due to the loss of stability of the meniscus \red{(see Fig.\ \ref{F_inestMeniscus}a)}, while the dripping observed on the left side of the vertical branch is produced by the breakup of the jet next to the discharge orifice \red{(see Fig.\ \ref{F_inestMeniscus}b)}. The horizontal branch of the stability limit is the so-called minimum flow rate stability limit, which is also characteristic of other tip streaming configurations such as gaseous flow focusing \citep{CHGM17} and electrospray \citep{PRHGM18}. This limit is very relevant at the technological level because it leads to the smallest droplets/emulsions for a given configuration.

The quantitative differences between the experimental and numerical results may be due to differences between the true and numerical geometries or the position of the triple contact line in both cases. In fact, while the feeding capillary end is assumed to be infinitely thin in the numerical simulation (Fig.\ \ref{numer2}), it has a noticeable thickness in the experiments. \blue{As mentioned in the Introduction,} the existence of unstable experimental realizations in the SJTS parameter region may be explained in terms of the growth of linear perturbations due to the short-term superposition of stable modes \citep{S07}. As occurs in gaseous flow focusing \citep{CHGM17}, this phenomenon can take place when the instability is localized in the emitted jet (convective instability branch). To assess the validity of this hypothesis, one has to examine the response of the system to small-amplitude perturbations from direct numerical simulations, which is beyond the scope of the present work.

In the rest of this section, we study numerically the stability of the liquid-liquid flow focusing configuration. Given the large dimension of the parameter space, a systematic stability analysis constitutes a formidable task. For this reason, we will restrict ourselves to the geometrical configuration considered in our experiments, and to the dimensionless numbers $\rho=1$ and Re=0.02, which are similar to those of the experiments as well. We will examine the effect of the viscosity and flow rate ratio, $\mu$ and $Q$, and the capillary number Ca on the SJTS linear stability.

\begin{figure}
\begin{center}
\resizebox{0.45\textwidth}{!}{\includegraphics{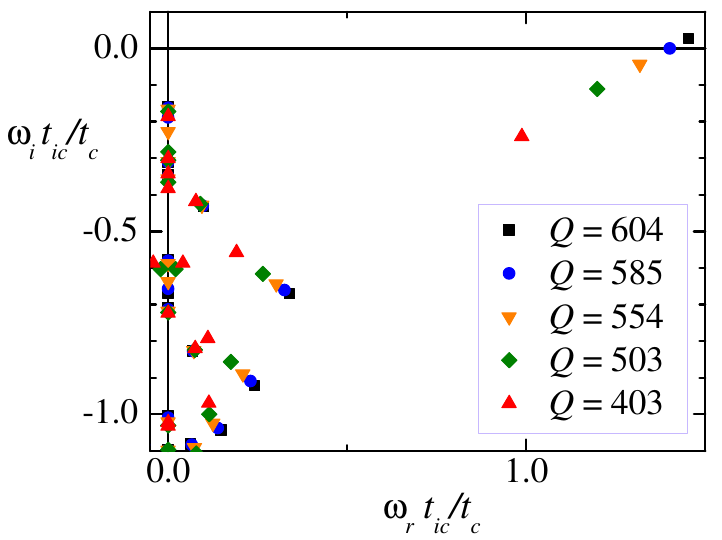}}
\end{center}
\caption{Spectrum of eigenvalues with $\omega_i t_{ic}/t_c>-1.1$ and $\omega_r t_{ic}/t_c\geq 0$ for $\rho=1$, Ca=0.05, Re=0.02, and $\mu=0.1$, and different flow rate ratios as indicated in the figure. Here, the eigenfrequency has been made dimensionless with the inertio-capillary time $t_{ic}=(\rho_i R_c^3/\sigma)^{1/2}$.}
\label{eigen}
\end{figure}

As explained in Sec.\ \ref{s4}, the stability of SJTS is assessed by calculating the global linear modes characterizing the response of the base flow to small-amplitude perturbations. For the sake of illustration, Fig.\ \ref{eigen} shows the spectrum of eigenvalues with $\omega_i t_{ic}/t_c>-1.1$ for $\mu=0.1$ and different flow rate ratios. Most of the eigenvalues have a real part within the interval $[-1/2,1/2]$. However, there is an eigenvalue whose real part lies outside that interval. The corresponding eigenmode becomes the dominant one as the flow rate ratio increases. This mode becomes unstable for $Q\simeq 585$, which is the critical flow rate ratio for this specific configuration. Figure \ref{eigen} illustrates the importance of conducting a careful analysis of the spectrum of eigenvalues, because the eigenmode responsible for instability is not necessarily the dominant one over the range of flow rates analyzed.  

\begin{figure}
\begin{center}
\resizebox{0.5\textwidth}{!}{\includegraphics{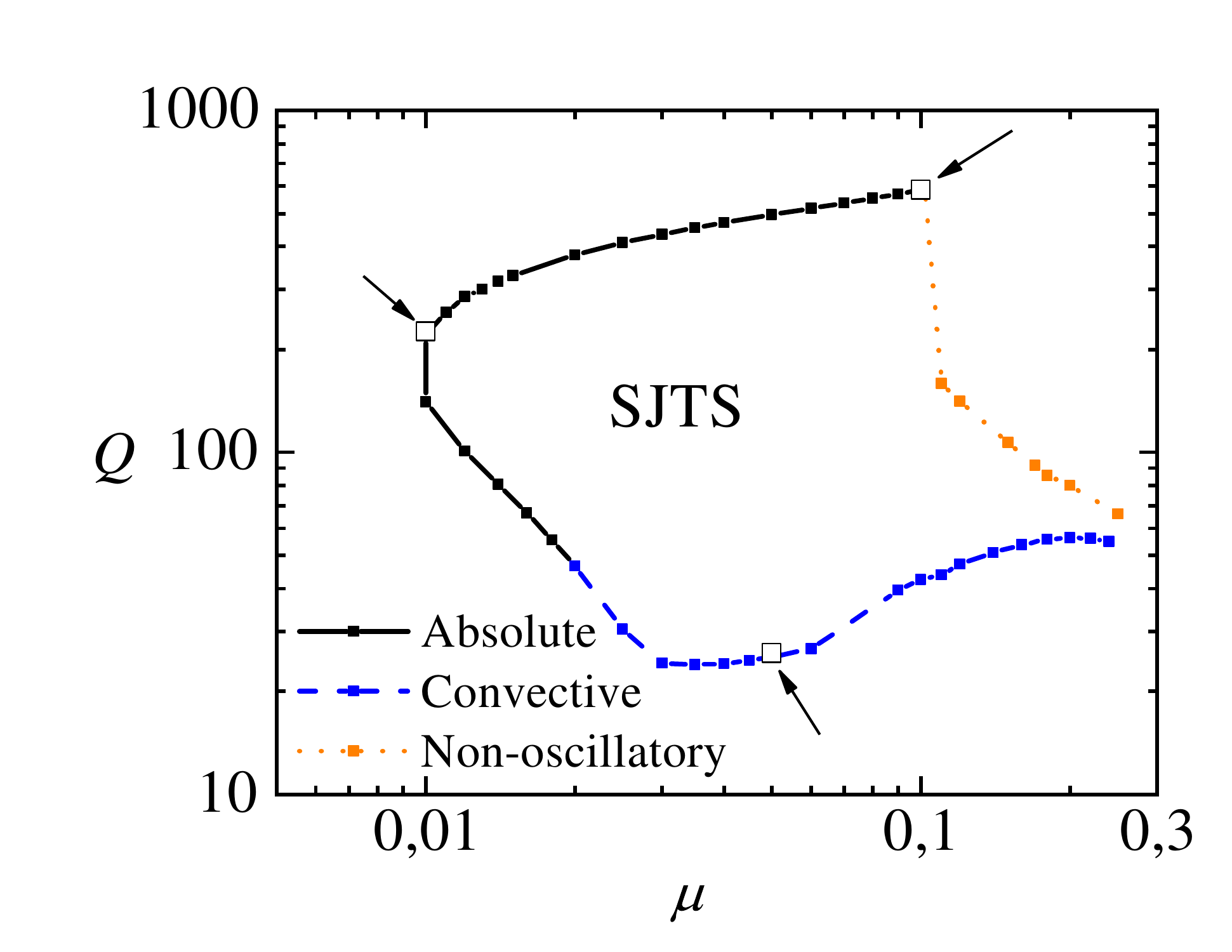}}
\end{center}
\caption{Stability island of SJTS in the ($\mu$,Q) parameter plane. The solid and dashed lines correspond to the absolute and convective instabilities, respectively. The dotted line corresponds to a non-oscillatory instability whose convective/absolute character cannot be determined. The arrows indicate the cases analyzed in Fig.\ \ref{modes}-\ref{forces}. The results were calculated for $\rho=1$, Ca=0.05, and Re=0.02.}
\label{numer}
\end{figure}

\begin{figure}
\begin{center}
\resizebox{0.75\textwidth}{!}{\includegraphics{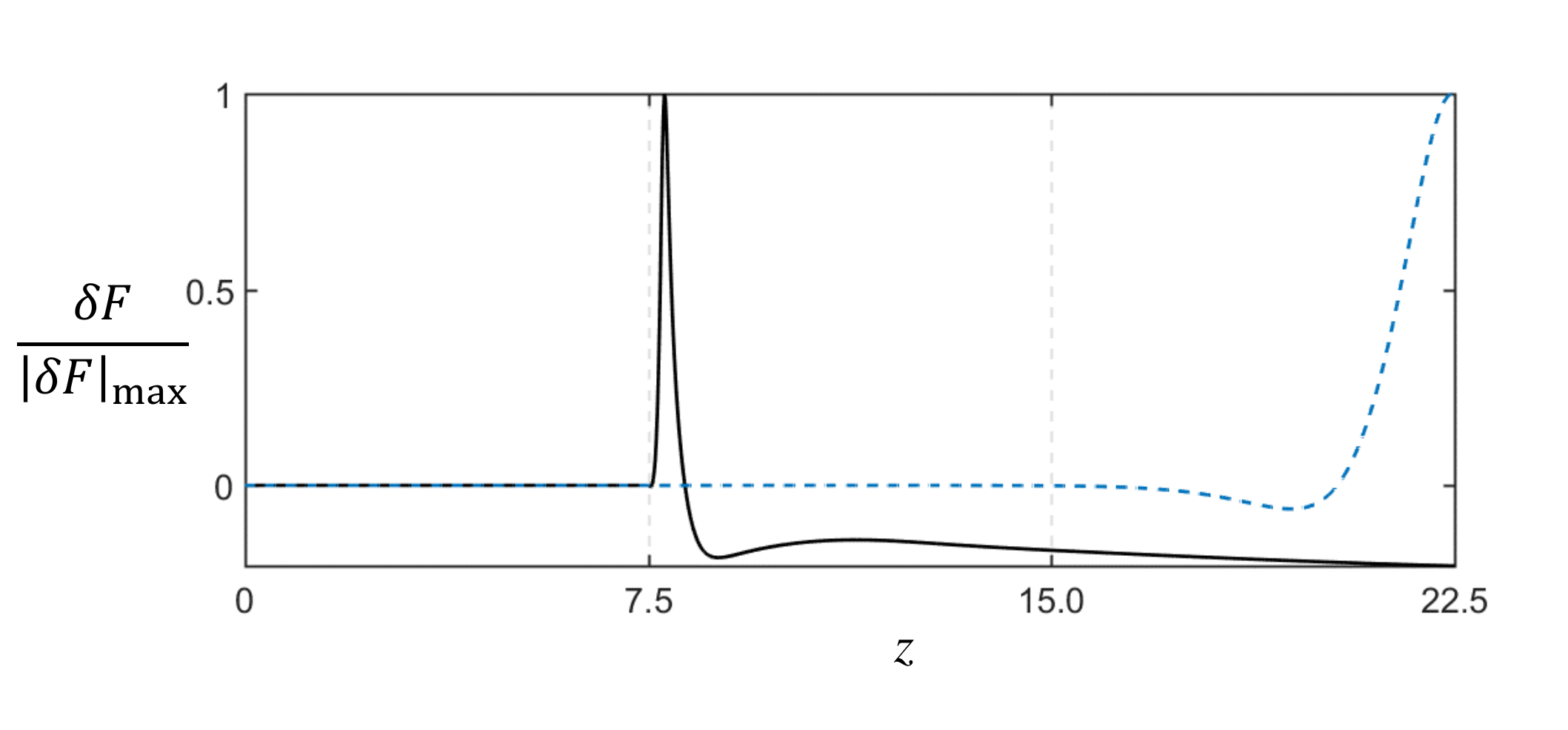}}
\end{center}
\caption{Amplitude of the interface perturbation, $\delta F$, normalized with the maximum value of its magnitude, $\left|\delta F\right|_{\textin{max}}$. The solid and dashed lines correspond to the cases ($\mu=0.1$, $Q=585$) and ($\mu=0.05$, $Q=30$) indicated with arrows in Fig.\ \ref{numer}, respectively. The results were calculated for $\rho=1$, Ca=0.05, and Re=0.02.}
\label{modes}
\end{figure}

Figure \ref{numer} shows the parameter region in the ($\mu$,Q) plane within which the base flow is linearly stable, and, therefore, SJTS can be obtained. As anticipated above, this stability island is limited by two types of instabilities: one associated with the destabilization of the tapering meniscus (absolute instability), and the other with the growth of capillary waves beyond the discharge orifice (convective instability). This distinction is made clear in Fig.\ \ref{modes}, where we show the amplitude of the interface perturbation for each of the two cases mentioned above. The solid line corresponds to a marginally stable flow in the absolute instability branch of Fig.\ \ref{numer}. In this case, the maximum of the perturbation amplitude is reached in the tapering meniscus, which indicates that the perturbation is originated in this region and then propagates throughout the rest of the liquid domain. On the contrary, the tapering meniscus remains practically stable in the base flow belonging to the convectively unstable branch (dashed line in Fig.\ \ref{modes}). In this case, the perturbation grows only beyond the discharge orifice leaving practically intact the tapering meniscus and part of the ejected liquid thread. There is a gradual transition from absolute to convective instability along the lower border of the stability island as $\mu$ increases. In other words, the maximum of the perturbation amplitude moves downstream as $\mu$ increases. This means that reducing the inner liquid viscosity stabilizes the tapering meniscus and destabilizes the emitted jet. It is worth mentioning that absolutely unstable base flows are always identified as dripping experimental realizations, while convective instability can correspond either to SJTS or to dripping depending on the distance of the interface breakup point from the discharge orifice. 

The \blue{small-amplitude perturbations} responsible for the destabilization of SJTS in Fig.\ \ref{numer} can be classified as oscillatory and non-oscillatory depending on whether the real part of the dominant mode eigenfrequency is different from or equal to zero, respectively. In the first case, the values of all the hydrodynamic quantities, including the interface position, oscillate while growing. On the contrary, the perturbation amplitude increases exponentially with time in the non-oscillatory case. This distinction cannot be easily established experimentally as the oscillation frequency typically scales as the inverse of the capillary time (see Fig.\ \ref{eigen}), and, therefore, the oscillations take place over a very short time. It is worth noting that non-oscillatory instabilities frequently manifest themselves as bifurcations of the base flow numerical solutions, and, therefore, they can be detected without conducting the linear stability analysis of those solutions. In fact, the non-oscillatory stability limit in Fig.\ \ref{numer} was determined as the parameter conditions for which the numerical method fails to converge to a proper base flow solution. For this reason, we could not establish its convective or absolute character. On the contrary, the base flow solution can surpass an oscillatory instability limit as one of the control parameters is varied without any qualitative change of its characteristics. This means that the global stability analysis is required to determine which of the calculated base flows truly correspond to an experimental realization. 

\blue{The oscillatory/non-oscillatory character of the small-amplitude perturbation should not be confused with the unsteady/quasi-steady character of the mode adopted by the flow after the destabilization of the steady jetting regime. In fact, non-oscillatory growing linear perturbations can lead to a pulsating mode in which an unsteady meniscus emits intermittently droplets as series of polydisperse trains, as occurs in the coflowing configuration \citep{GSC14}. On the contrary, oscillatory small-amplitude perturbations produce the transition to a dripping mode in which large droplets are ejected by a quasi-steady meniscus next to the discharge orifice.}

The smallness of the Weber number, We$= \rho_i v_c^2/(\sigma/R_c)$=Ca Re=0.001, reveals the importance of the focusing effect to produce steady jetting in the present configuration. The action of the outer stream allows for the ejection of a liquid thread even though liquid inertia is much smaller than the resistant capillary force in the feeding tube. The results in Fig.\ \ref{numer} show that SJTS cannot be obtained for $Q\lesssim 20$, which indicates that the axial momentum transferred by the outer stream must be sufficiently large to focus the inner liquid current. The absolute instability arising when $Q$ exceeds a critical value for a given viscosity ratio corresponds to the classical minimum (inner) flow rate stability limit. As mentioned above, this stability limit is particularly important at the technological level because it leads to the smallest droplets produced with that microfluidic configuration. 

The physical mechanism responsible for the minimum flow rate stability limit in SJTS is still a matter of debate. For gaseous flow focusing \citep{G98a}, it has been speculated that the loss of stability of low-viscosity menisci is caused by the growth of recirculation cells \citep{MRHG11}. In fact, the destruction of those cells by modifying the emitter geometry has an important stabilizing effect not only in gaseous flow focusing \citep{ARMGV13} but also in the cone-jet mode of electrospray \citep{MRRS16}. In gaseous flow focusing, the viscosity of the focused liquid stream ``arranges the flow pattern"\, in the tapering meniscus, which prevents the formation of recirculation cells. This explains the significant stabilizing effect of viscosity in that configuration \citep{MRHG11}. On the contrary, Fig.\ \ref{numer} shows that the minimum flow rate of the dispersed phase increases as the viscosity of that phase increases ($Q$ decreases with $\mu$), which means that the inner viscosity has a destabilizing effect in the liquid-liquid configuration. 

\begin{figure}
\begin{center}
\resizebox{0.55\textwidth}{!}{\includegraphics{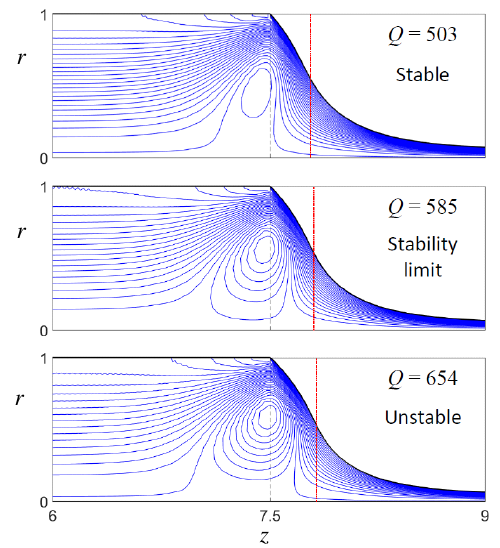}} 
\end{center}
\caption{Streamlines in the inner liquid stream for $\mu=0.1$ and the flow rates indicated in the figure. The red vertical line indicates the location of maximum amplitude of the interface perturbation. The results were calculated for $\rho=1$, Ca=0.05, and Re=0.02.}
\label{cell}
\end{figure}

\begin{figure}
\begin{center}
\resizebox{0.5\textwidth}{!}{\includegraphics{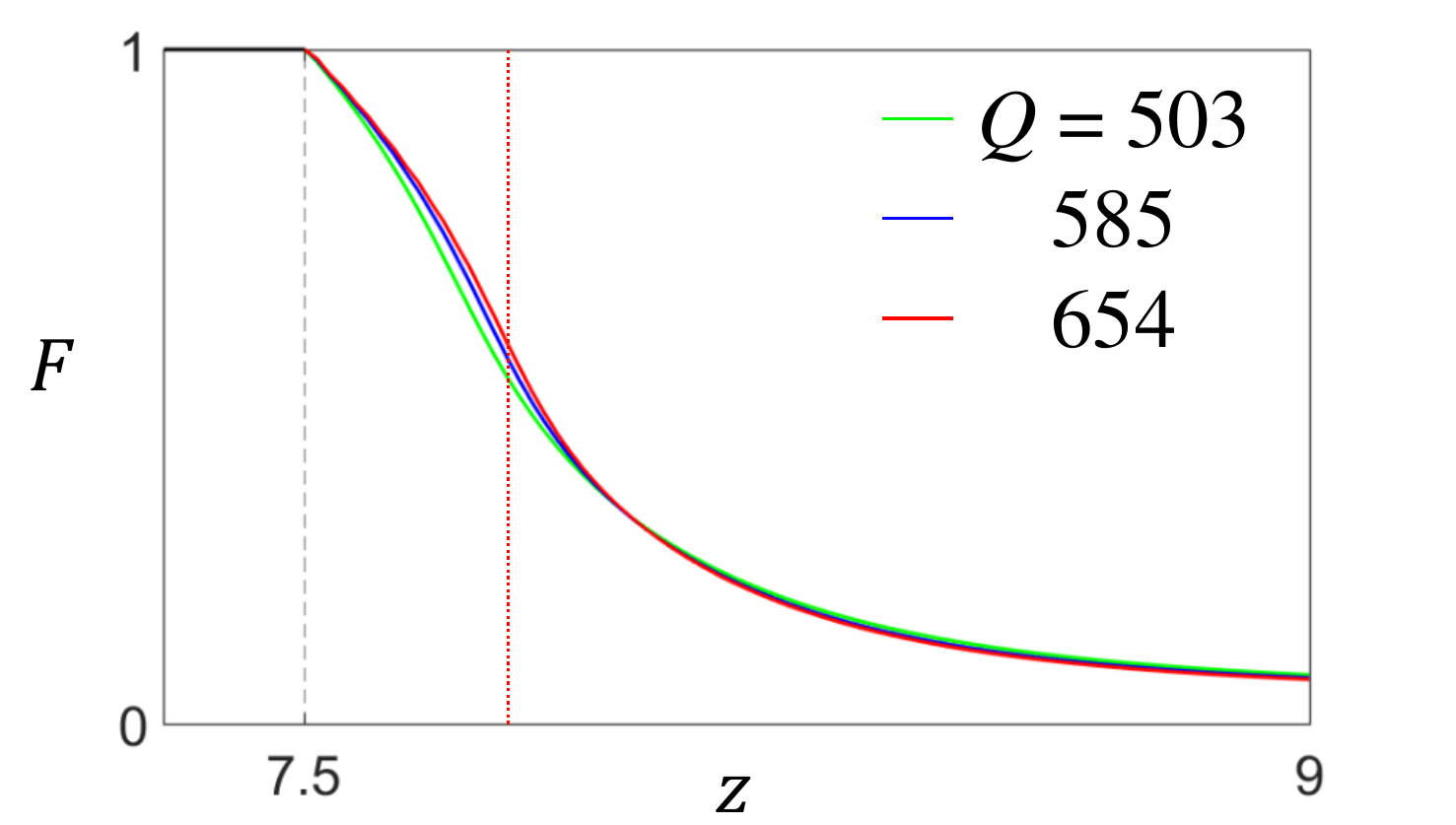}} 
\end{center}
\caption{Liquid-liquid interface for $\mu=0.1$ and the flow rates indicated in the figure. The red vertical line indicates the location of maximum amplitude of the interface perturbation at the stability limit. The results were calculated for $\rho=1$, Ca=0.05, and Re=0.02.}
\label{defor}
\end{figure}

We examine in Fig.\ \ref{cell} the streamlines in base flows next to the (inner) minimum flow rate stability limit (maximum flow rate ratio). Owing to the relatively small value of the inner viscosity (relative large value of the ratio $\mu$), a recirculation cell is formed right at the exit of the feeding capillary. As the flow rate ratio approaches its maximum value $Q\simeq 585$, the cell gets closer to the interface narrowing the fluid passage through which the ejected liquid crosses the meniscus. As this passage narrows, the interface slightly deforms and an inflection point appears in the interface contour (Fig.\ \ref{defor}). At the stability limit, the interface does not withstand perturbations of the hydrodynamic fields, which produce oscillations of the interface that grow on time leading to the base flow destabilization. In fact, the maximum of the interface perturbation amplitude (Fig.\ \ref{modes}) is located very close to the inflection point of the interface contour. As will be shown below, a small reduction of the Capillary number (an increase of the interfacial tension) results in a decrease of the minimum inner flow rate of the same order of magnitude. In other words, the ``hardening"\ of the interface significantly stabilizes the base flow at the minimum flow rate stability limit.

The vertex ($\mu\simeq 0.1$, $Q\simeq 600$) of the stability border in Fig.\ \ref{numer} corresponds to the appearance of a non-oscillatory dominant mode when $\mu$ exceeds the critical value $\mu\simeq 0.1$. The physical mechanism associated with this mode is expected to be substantially different from that of the oscillatory instability described above. \blue{The hydrostatic pressure and viscous stress normal to the interface deform the meniscus surface. This deformation is withstood by the interfacial tension. For the parameter conditions corresponding to this stability limit, the interfacial tension no longer balances the other normal stresses, and SJTS becomes unstable}. In fact, we have verified that small relative variations of the interfacial tension (Capillary number) produce relative variations of the same order of magnitude of the location of the vertex ($\mu\simeq 0.1$, $Q\simeq 600$). \blue{A similar instability occurs in, e.g., the minimum volume stability limit of liquid bridges \citep{VMHF14}, in which the surface tension cannot withstand the axisymmetric liquid bridge deformation caused by the volume reduction.}

\begin{figure}
\begin{center}
\resizebox{0.65\textwidth}{!}{\includegraphics{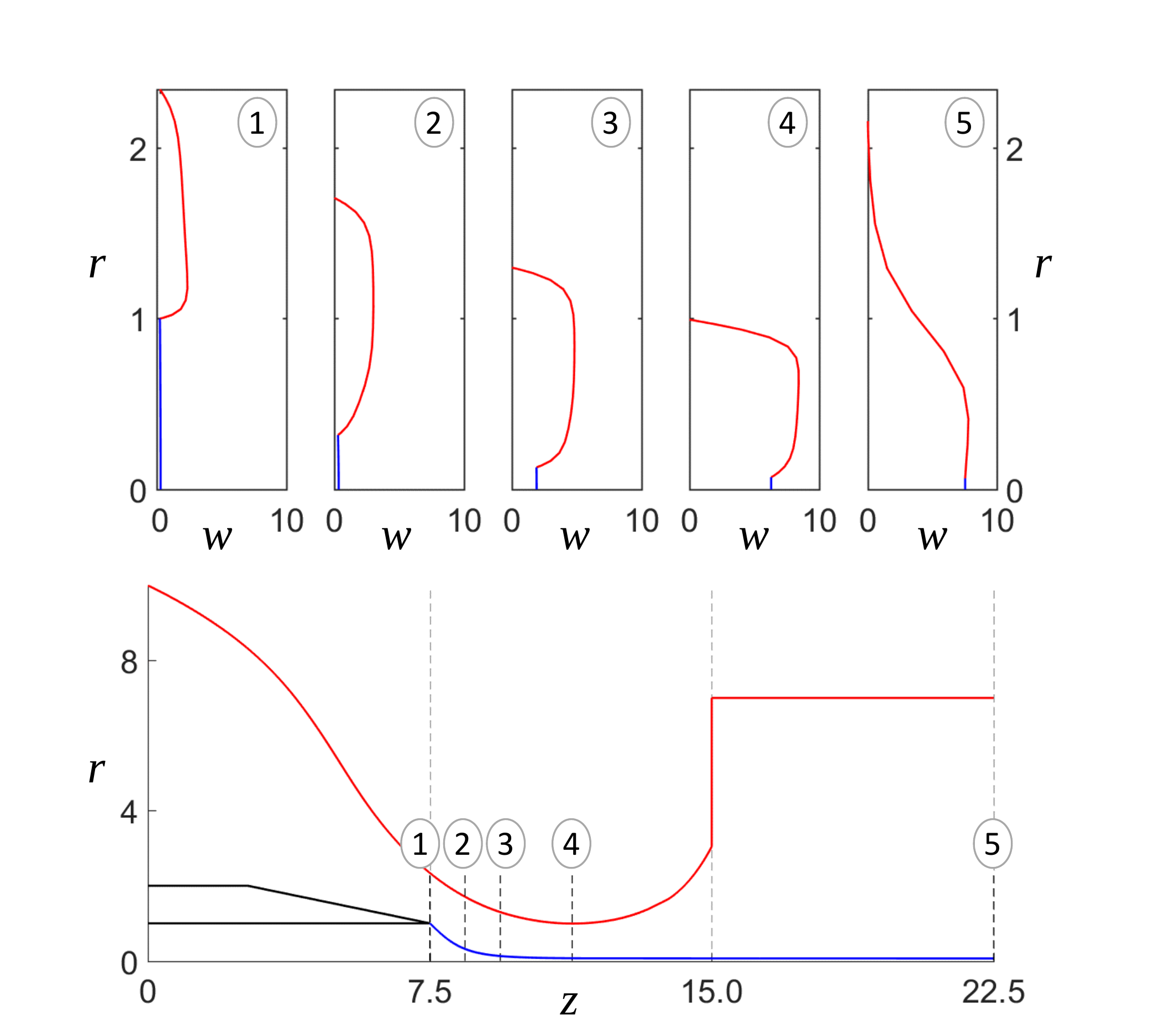}}
\end{center}
\caption{Axial velocity profile at several cross sections for $\mu=0.01$ and $Q=225$. The results were calculated for $\rho=1$, Ca=0.05, and Re=0.02.}
\label{velocity}
\end{figure}

\begin{figure}
\begin{center}
\resizebox{0.65\textwidth}{!}{\includegraphics{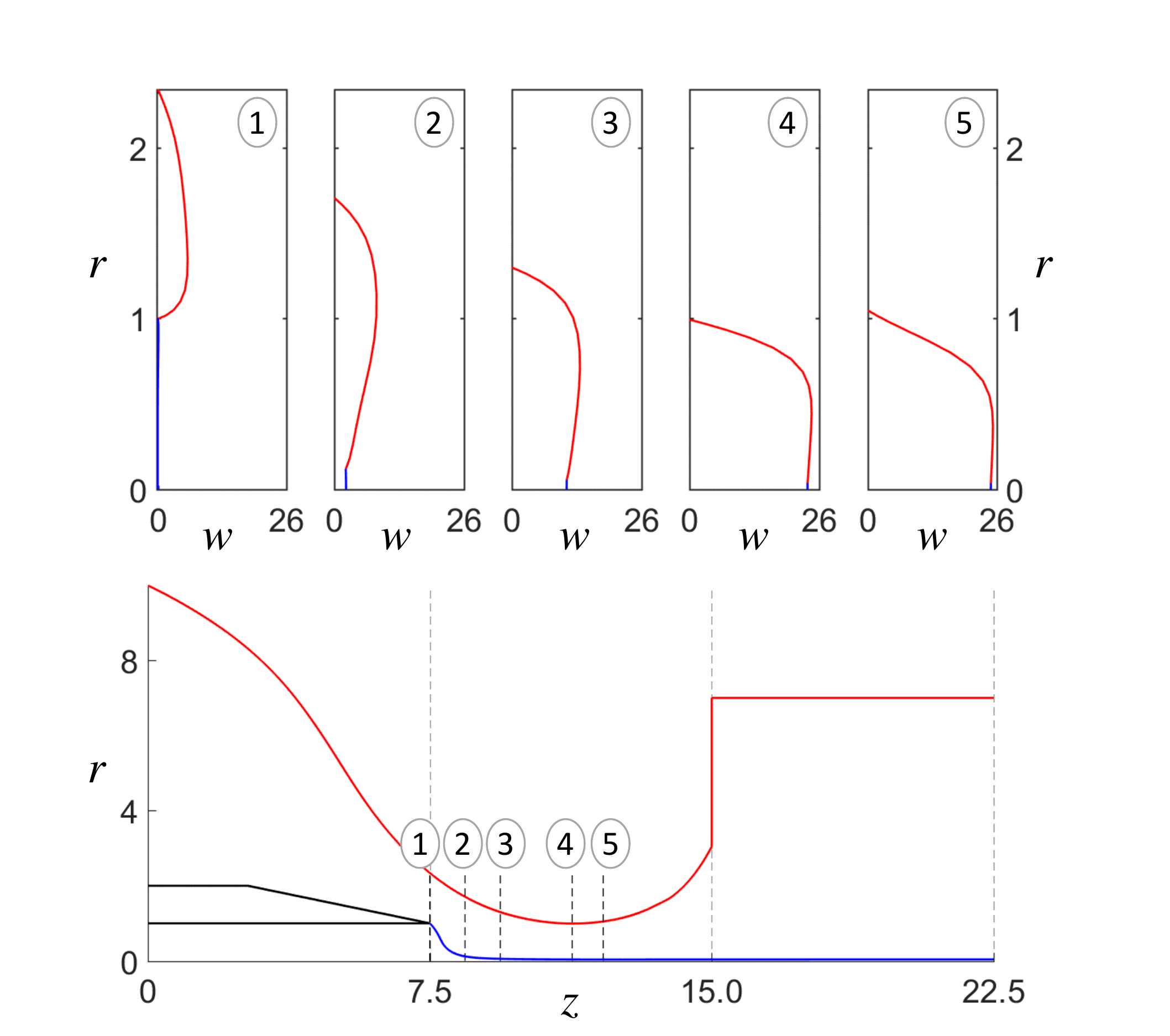}}
\end{center}
\caption{Axial velocity profile at several cross sections for $\mu=0.1$ and $Q=585$. The results were calculated for $\rho=1$, Ca=0.05, and Re=0.02.}
\label{velocity2}
\end{figure}

Figure \ref{velocity} shows the velocity profile at several cross-sections of the marginally stable base flow ($\mu=0.01$, $Q=225$). The boundary layer grows at the inner wall of the nozzle. Mass conservation makes the inviscid core of the outer stream accelerate in the nozzle. Axial momentum is transferred to the inner liquid jet due to the collaboration of the hydrostatic pressure drop in front of the nozzle neck and the shear viscous stress at the interface. The maximum velocity of the outer stream in the discharge bath is not reached at the interface but a considerable distance from it. For this reason, viscous stress transfers axial momentum from the outer stream to the inner current beyond the discharge orifice, which causes an extra acceleration of the jet in that region. When the viscosity ratio increases (Fig.\ \ref{velocity2}), the outer stream transfers faster its axial momentum to the jet, and the latter reaches its maximum speed inside the nozzle. 

\begin{figure}
\begin{center}
\resizebox{0.5\textwidth}{!}{\includegraphics{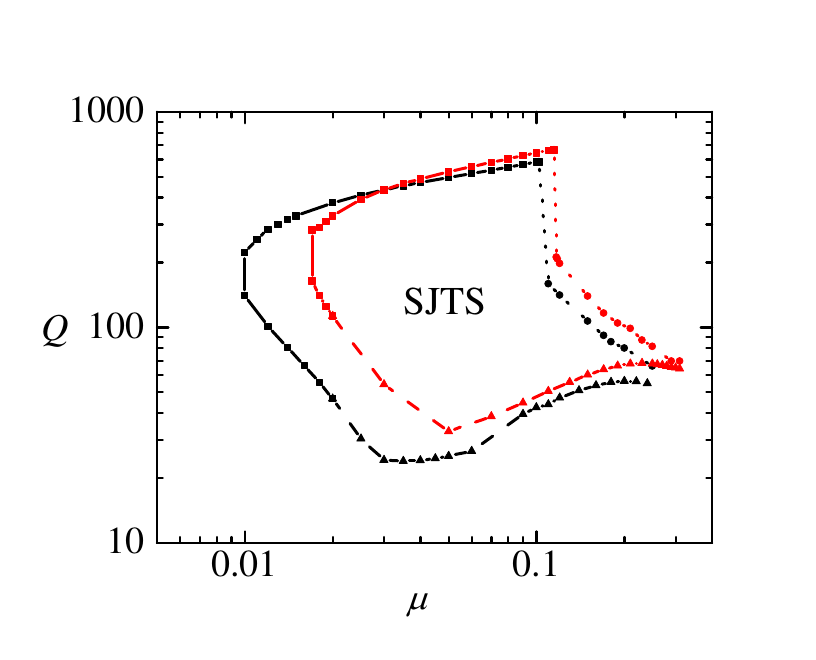}}
\end{center}
\caption{\red{Stability island of SJTS in the ($\mu$,Q) parameter plane calculated for $\rho=1$, Re=0.02, and Ca=0.05 (black lines and symbols) and 0.0375 (red lines and symbols). The solid and dashed lines correspond to the absolute and convective instabilities, respectively. The dotted line corresponds to a non-oscillatory instability whose convective/absolute character cannot be determined.}}
\label{numer3}
\end{figure}

\red{Figure \ref{numer3} shows how the capillary number affects the stability region where the base flow is linearly stable. Decreasing the capillary number has a destabilizing effect on the left and lower borders of the island, showing that the axial momentum necessary to produce the focusing effect and to achieve SJTS becomes significantly larger. However, the reduction of the capillary number has a stabilizing effect at the upper and right boundaries, which are more interesting from the technological point of view. The interface can withstand larger stresses due to the increase of the interfacial tension. Therefore, the minimum inner flow rate decreases ($Q$ increases), while the viscosity ratio corresponding to the upper-right vertex of the stability island ($\mu=0.115$, $Q=665$) increases.}

\begin{figure}
\begin{center}
\resizebox{0.65\textwidth}{!}{\includegraphics{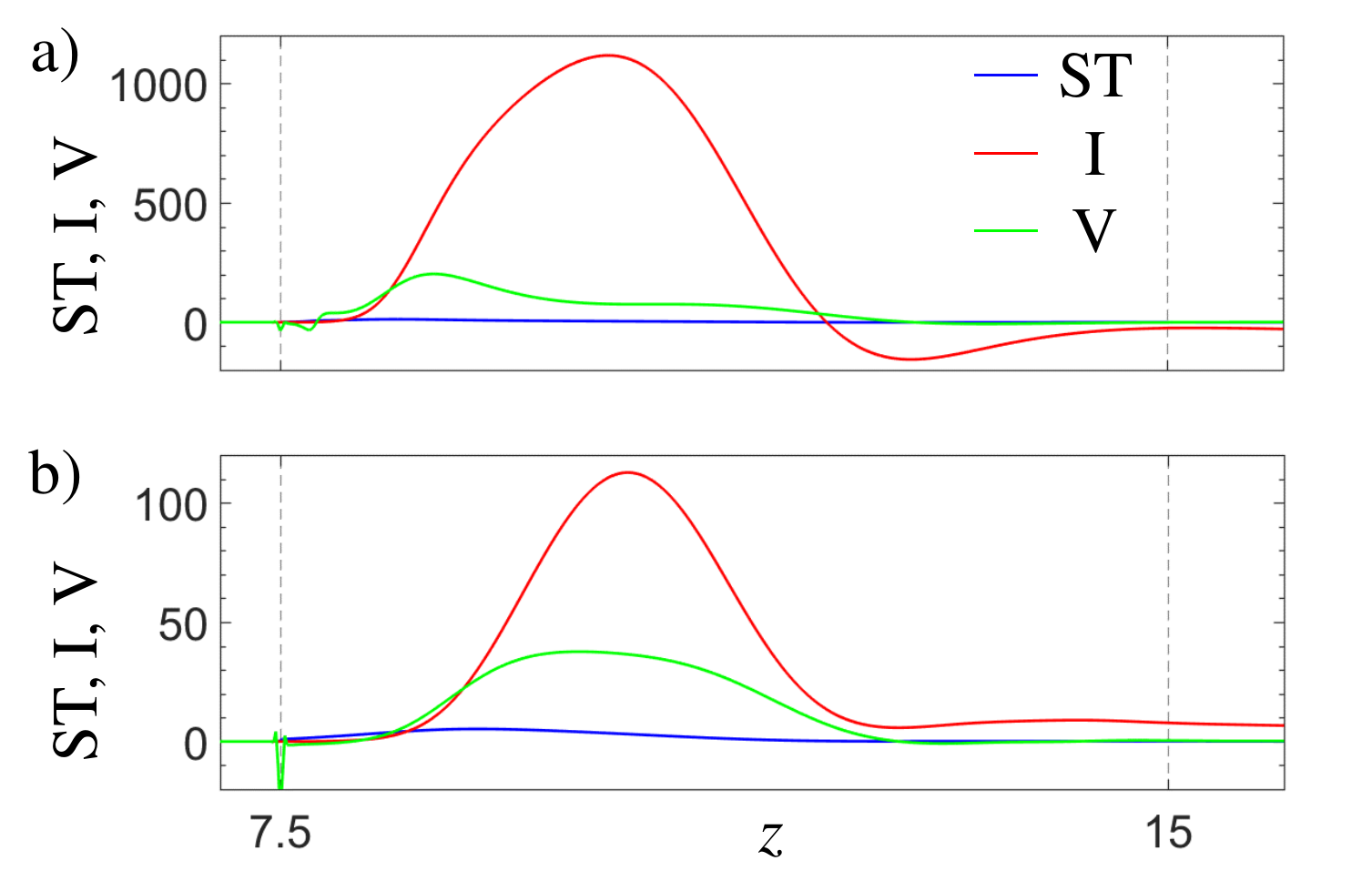}}
\end{center}
\caption{Axial forces per unit volume due to surface tension (ST), inertia (I), and viscosity (V) for a) ($\mu=0.1$, $Q=585$) and b) ($\mu=0.01$, $Q=225$). The dashed lines indicate the needle tip and the nozzle exit positions. The results were calculated for $\rho=1$, Ca=0.05, and Re=0.02.}
\label{forces}
\end{figure}

The 1D (slenderness) approximation is a simple and useful way of quantifying the forces arising in the inner liquid jet throughout its emission. \blue{This approximation is derived by considering the radial Taylor expansion of the hydrodynamic fields \citep{E97}, and is valid when the inner fluid adopts a slender shape along the streamwise direction}. In this model, the momentum equation in the $z$-direction becomes \citep{MG20}:
\begin{eqnarray}
\label{mom}
{\cal F}=
\underbrace{\left(\frac{1}{F}\right)_z}_{\text{\normalsize ST}}+
\underbrace{\left(\frac{Q^2 }{2F^4}\right)_z}_{\text{\normalsize I}}+
\underbrace{\frac{6\text{Oh} Q}{F^2}\left(\frac{F_z}{F}\right)_z}_{\text{\normalsize V}}.\nonumber\\
\end{eqnarray}
The term ${\cal F}$ comprises the hydrostatic pressure and viscosity forces per unit volume exerted by the outer stream on a slice of the inner liquid between $z$ and $z+dz$. The terms on the right-hand side of Eq.\ (\ref{mom}) are the resistant forces due to surface tension (ST), inertia (I), and viscosity (V), respectively. Figure \ref{forces} shows the values taken by these three terms for ($\mu=0.1$, $Q=585$) and ($\mu=0.01$, $Q=225$). The area enclosed by the curves equals the work done/energy consumed per unit volume by the corresponding term. The 1D approximation is expected to give accurate results in the meniscus-to-jet transition and in the rest of the jet. As can be observed, surface tension is subdominant in the two cases considered. Most of the work done by the outer stream transforms into kinetic energy for the lower viscosity case $\mu=0.1$. When the inner viscosity increases ($\mu=0.01$), the resistant viscous force becomes a more significant sink of energy in the central part of the nozzle. For $\mu=0.1$, the minimum radius is reached inside the nozzle, which explains the change of the sign of the inertia term. 

\begin{figure}
\begin{center}
\resizebox{0.5\textwidth}{!}{\includegraphics{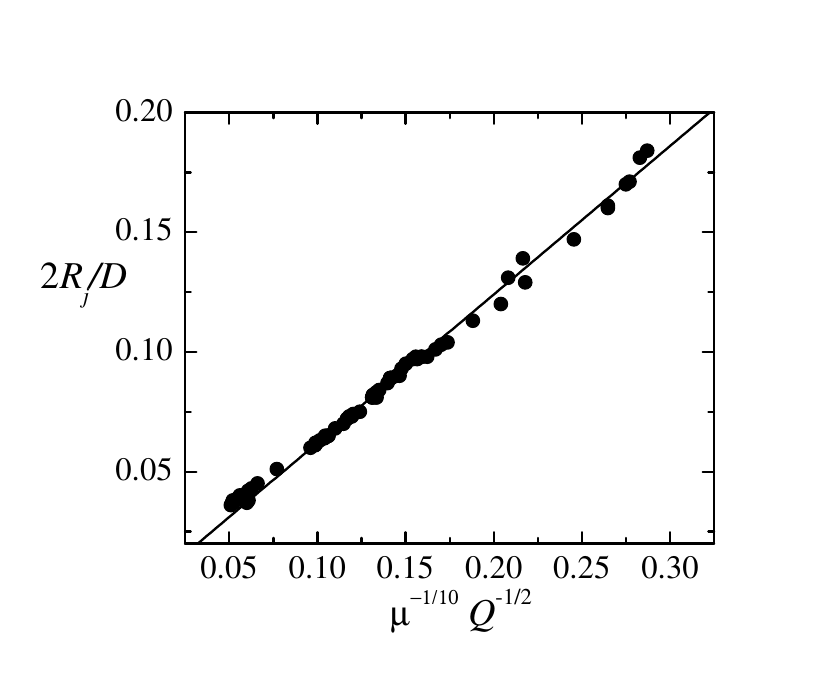}}
\end{center}
\caption{Radius of the emitted jet calculated from the simulations. The solid line is the scaling law (\ref{sc}).}
\label{scaling}
\end{figure}

Boundary layers grow on the nozzle inner wall due to viscous stresses in the outer stream. In addition, the jet and outer streams move at the same speed in the nozzle neck for $Q^{-1}\ll 1$ ($Q_i\ll Q_o$). Therefore, if one neglects viscous effects in the outer stream, then mass conservation yields $2R_j/D=Q^{-1/2}$ for $Q^{-1}\ll 1$. As described above, the growth of the boundary layers accelerates the inviscid core of the outer stream for a fixed outer flow rate, which enhances the transfer of axial momentum to the inner liquid jet. This reduces the jet radius with respect to that predicted by the above inviscid approximation. Using the optimization method described by \citet{MG20}, the best collapse of the data around the scaling law $R_j/D\sim \mu^{\alpha} Q^{\beta}$ is obtained for $\alpha\simeq -0.05$ and $\beta\simeq -0.45$. Figure \ref{scaling} shows that
\begin{equation}
\label{sc}
\frac{2R_j}{D}=0.62\, \mu^{-1/10} Q^{-1/2}    
\end{equation}
constitutes a good approximation for $\rho=1$, Ca=0.05, and Re=0.02. The minimum value of $2R_j/D$ was 0.038, which shows the strong focusing effect achieved with this configuration. \red{Unfortunately, we could not take sufficiently sharp images of the emitted jets to measure accurately their diameters in most experimental realizations due to the presence of droplets in the outer bath. Nevertheless, Fig.\ \ref{valido} shows the remarkable agreement between the experimental and numerical interface contours. For this reason, we expect the experimental diameters to fit the scaling law (\ref{sc}) as well.}

\section{Conclusions}

We have studied the stability of SJTS produced with axisymmetric liquid-liquid flow focusing. For this purpose, both the base flow and its linear eigenmodes have been calculated numerically for arbitrarily values of the governing parameters. The critical conditions at which linear instability appears have been determined in the ($\mu$,$Q$) parameter plane for given values of $\rho$, Ca, and Re. The results show the existence of a parameter island within which SJTS is stable under small-amplitude perturbations. This island is delimited by both oscillatory and non-oscillatory instabilities. The unstable perturbations can be originated in the tapering meniscus (absolute instability) or beyond the discharge orifice (convective instability) depending on the values of the viscosity and flow rate ratios. We have examined the stability limit corresponding to the minimum inner flow rate, which leads to the production of the smallest droplets. For small inner viscosity, this instability may be caused by the displacement towards the interface of the recirculation cell formed in the tapering meniscus, which narrows the stream tube across which the injected liquid leaves the meniscus. We have verified that a decrease of the Capillary number significantly shifts the stability island shown in Fig.\ \ref{numer} both upwards and rightwards. We have calculated the jet diameter for the marginally stable flows, and have derived a simple scaling law for that quantity. This law is based on mass conservation slightly corrected to account for the viscosity forces. We have conducted experiments to validate our numerical approach. \red{The experiments confirm the existence of the two instability mechanisms predicted by the global stability analysis.} 

The present analysis has potential applications in the development of new microfluidic techniques to produce microemulsions with a high degree of monodispersity and sizes much smaller than that of the feeding capillaries. A possibility is to adapt the stabilization techniques applied to gaseous flow focusing or electrospray to the present liquid-liquid configuration. Thus, a pointed bar placed coaxially with the feeding capillary would destroy the recirculation cell formed in low-viscosity tapering menisci. A natural question is whether this effect may reduce the minimum inner flow rate, and, therefore, the minimum achievable size of the produced droplets.

\vspace{1cm}
{\bf Acknowledgement.} This research has been supported by the Spanish Ministry of Economy, Industry and Competitiveness under Grant DPI2016-78887, by Junta de Extremadura under Grant GR18175 and by Junta de Andalucía under Grant P18-FR-3623.


\end{document}